\documentclass[12pt]{article}
\usepackage{epic}
\usepackage{latexsym}

\renewcommand{\thesection}{\Roman{section}}
\newtheorem{theorem}{Theorem}

\newtheorem{lem}{Lemma}
\renewcommand{\appendix}{%
\newcommand{\Section}{%
\secdef\Appendix\sAppendix}%
\setcounter{section}{0}%
\renewcommand{\thesection}{\Alph{section}}%
}

\newcommand{\Appendix}[2][?]{%
\refstepcounter{section}%
\addcontentsline{toc}{appendix}%
{\protect\numberline{\appendixname~\thesection} #1}%
{\flushleft\large\bfseries\appendixname\ \thesection\par
#2
\par}%
\sectionmark{#1}\vspace{\baselineskip}%
\setcounter{equation}{0}%
\renewcommand{\theequation}{\thesection\arabic{equation}}%
}

\newcommand{\sAppendix}[1]{
{\flushleft\large\bfseries\appendixname\par
#1\par}
\vspace{\baselineskip}}
\begin{document}


\vskip 0.7truecm

\begin{center}
{ \bf ON PATH INTEGRAL LOCALIZATION AND THE LAPLACIAN \\  }
\end{center}

\vskip 1.5cm

\begin{center}
{\bf Topi K\"arki $^{*}$ } \\
\vskip 0.4cm
{\it Institute Mittag-Leffler, 
Auravagen 17, S-18262, Djursholm, Sweden}
\vskip 0.4cm
\end{center}

\vskip 2.7cm

\rm
\noindent
We introduce a new localization principle
which is a generalized canonical transformation. It unifies  
BRST localization, the non-abelian localization principle
and a special case of the conformal Duistermaat-Heckman integration formula of
Paniak, Semenoff and Szabo. The heat kernel on compact Lie groups
is localized in two ways. First using a non-abelian
generalization of the derivative expansion localization of
Palo and Niemi and secondly using the BRST localization principle
and a configuration space path integral.
In addition we present some new formulas on homogeneous spaces
which might be useful in a possible localization of Selberg's trace formula on locally homogeneous spaces.
\vfill
31.15.K, 02.40.H, 02.20.T
\begin{flushleft}
\rule{5.1 in}{.007 in}\\
$^{*}$ {\small E-mail: karki@ml.kva.se \\}
\end{flushleft}

\newpage


\section{Introduction}
Integral localization is a method to calculate 
path integrals. It conventionally involves a BRST
symmetry and a one-parameter localization deformation of the action
which in the limit that the localization parameter is put
to infinity evaluates the integral\cite{thesispalo}\cite{thesistirkkonen}\cite{locdiag}. 
The result is usually a sum over the critical points
of the action or an integral over a finite dimensional subspace of
the original infinite dimensional integration domain.
The method
has been effective in topological field theories\cite{topofield}.

A special case of it is the
Duistermaat-Heckman integration formula\cite{own} and its loop space version\cite{wkb} where
the BRST symmetry is another way of writing the definition 
of the Hamiltonian vector field. A more technical introduction
to the Hamiltonian BRST symmetry is given in Sec. \ref{brs}. 

Recently there has emerged localizations
that cannot be understood from the conventional BRST point of view: 
non-abelian localization principle\cite{2qc} and 
the conformal Duistermaat-Heckman
formula of Paniak, Semenoff and Szabo\cite{con}.
In Secs. \ref{new}, \ref{non} and \ref{con:sec} they are
unified in a new localization principle.

The two main questions behind this article are ``Why the
heat kernel of the Laplacian on compact Lie groups is
semiclassically exact\cite{mar}\cite{pic}?'' and
``Why there are Selberg's trace formulae\cite{gut}\cite{hej}\cite{bus}\cite{gro}?''. The
new localization principle was also discovered in studying these questions.
We review shortly the facts that motivated them. 
The heat kernel on Lie groups has been proved to be semiclassically
exact\cite{mar}\cite{pic} by direct comparison. On the other hand
the loop space Duistermaat-Heckman theorem\cite{wkb} explains
semiclassical exactness, in the case that the Hamiltonian
vector field is also a Killing vector field, using a path integral localization
proof. It is not however the case (in the obvious way) on
Lie groups.
In addition there is Selberg's trace formula on constant negative
curvature Riemann surfaces 
\begin{equation}
tr\:e^{-\beta \triangle_0}=F
+(4\pi \beta)^{-\frac{1}{2}}e^{-\beta/4}\sum_{n=1}^{\infty}\sum_{p}
\frac{l(p)/2}{\sin l(p^n)/2}e^{-l^2(p^n)/4\beta} \label{selberg} 
\end{equation}
where $F$ is the fixed point contribution\cite{bus}. $p$ means a
primitive geodesic and $l(p)$ the length of it. The formula
is similar to the semiclassical approximation because it holds that
$S=\frac{1}{4}\int_0^\beta g_{\mu\nu}\dot{x}^{\mu}\dot{x}^{\nu}=l^2(p^n)/4\beta$
at the critical points of the action $S.$ The negative constant curvature
Riemann surfaces
can be obtained as quotient spaces $\Gamma\backslash(SU(1,1)/U(1))$ where $\Gamma$
is a discrete subgroup\cite{hej}. This
together with the semiclassical exactness on Lie groups
and the fact that there
are also generalizations of Selberg's
trace formula on some symmetric spaces\cite{gro} leads to the speculation of a localization
of the Laplacian on locally homogeneous manifolds $\Gamma\backslash (G/H).$

The answers that we provide to the questions above are organized as follows.

A localization deformation of the Laplacian on homogeneous spaces
is given in Sec. \ref{sec:lap}. The formulas and notations are presented in Appendix \ref{sec:geo} and \ref{sec:pha} because
most of them are not needed in the rest of the discussion. 
They involve some
new formulas: the use of a degenerate basis for vector fields is new as well as the Maurer-Cartan connection on the
tangent bundle and the phase space metric. In Appendix \ref{appb}
the scalar curvature is calculated. In
addition a formula is presented which might appear
in the hypothetical localization on locally homogeneous manifolds.

The question about the heat kernel on Lie groups is 
answered partially in Sec. \ref{ano} using a non-abelian derivative expansion localization. The localization principle is even more interesting
because it formally seems to apply 
to homogeneous manifolds as well, and because it may even apply to integrable models
(Sec. \ref{sec:conc}).
The final answer to the question is given in Sec. \ref{conf}
using a configuration space path integral and the symplectic form pointed 
out by R. Picken in Ref. \cite{pic}.

The question about Selberg's trace formulae is left open
despite the negative result, that it
is not the deformation in Sec. \ref{sec:lap}, 
and the speculations in Sec. \ref{sec:conc}.

In addition a new bound for the geodesic action is
presented in Appendix \ref{appa}.

\section{BRST localization principle}\label{brs}
We review the (Hamiltonian) BRST localization
principle introducing some notations.

We assume that $\Gamma$ is a $2D$-dimensional phase space,
$\omega$ a symplectic form and $H$ a Hamiltonian.
The classical partition function of the Hamiltonian system is
\begin{equation}
Z=\int_{\Gamma}\omega^De^{-\beta H},\label{partitionfunction}
\end{equation}
and it is calculated using  the BRST localization principle.
The definition of the Hamiltonian vector field $\chi$ can be written
as
\begin{equation}
(d+i_{\chi})(H+\omega)=0
\end{equation}
where $d$ is the exterior derivative and $i_{\chi}$ is a contraction 
operator. The equivariant derivative
$$d+i_{\chi}=d_{\chi}=Q$$
is a BRST symmetry with the exception that it is
closed only in the invariant subspace
$$Q^2\alpha=L_{\chi}\alpha=0$$
where $L_{\chi}$ is the Lie derivative and $\alpha$ is a differential form
on the phase space.
The BRST localization principle can be formulated as follows:
Analogously to the Fradkin-Vilkovisky theorem, 
the partition function (see the explanation of notations below)
\begin{equation}
Z_{\lambda}=\int d\phi^{\mu}\:d\psi^{\mu}e^{-\beta(H+\omega)+\lambda 
d_{\chi}\psi}\label{partitionfunction2}
\end{equation}
is independent of $\lambda$ provided that the arbitrary one-form 
$\psi$ satisfies the Lie derivative condition
\begin{equation}
L_{\chi}\psi=0.
\end{equation}
In (\ref{partitionfunction2})
$\phi^{\mu}$ are the coordinates on the phase space $\Gamma$
and the Grassmann variables $\psi^{\mu}$ are associated with the
one-forms $d\phi^{\mu}$ on the phase space. In particular the symplectic
form has been replaced by a bilinear in the fermionic variables $\omega_{\mu\nu}
d\phi^{\mu}\wedge d\phi^{\nu}\rightarrow\omega_{\mu\nu}
\psi^{\mu}\psi^{\nu}$. When $\lambda$ vanishes
the integral reduces to 
the partition function (\ref{partitionfunction}) modulo
an irrelevant factor\footnote{We use throughout the paper the convention
that such normalization factors are neglected which is usual
with path integrals.}
 of $(-\beta)^D$.
If one makes a clever choice of the one-form $\psi$ and takes the
localization parameter $\lambda$ to infinity one is able to evaluate the integral.

One can also give a loop space version
of the localization principle. The path integral\footnote{
We use the convention that the imaginary unit i is missing.} 
describing the quantum partition function is
\begin{equation}
Z_{\lambda}(\beta)=\int [d\phi^{\mu}\:d\psi^{\mu}]\exp {\int_0^{\beta}\theta_{\mu}
\dot{\phi}^{\mu}-H+\omega+\lambda d_S \psi}
\end{equation}
where $\omega$ is the loop space symplectic form, $\theta$ is the 
symplectic potential and $d_S$ is the loop space equivariant 
derivative 
\begin{equation}
d_S=d+i_{\dot{\phi}-\chi}=d+i_{\chi_S}.
\end{equation}
In addition $\chi_S$ is called the loop space Hamiltonian vector field and 
$\dot{\phi}=\dot{\phi}^{\mu}\frac{\partial}{\partial \phi^{\mu}}\equiv
\int_0^{\beta}dt\,\dot{\phi}^{\mu}(t)\frac{\delta}{\delta\phi^{\mu}(t)}$ is
a vector field on the loop space. We use the notation
that the integral sign is not written and the functional
derivatives are written as ordinary derivatives\cite{wkb}.
The loop space one-form $\psi$ must also satisfy the Lie derivative condition
$$L_{\chi_S}\psi=0.$$
Various applications of the principle can be found in Refs. \cite{wkb}, \cite{tir}, \cite{witloc} and \cite{witloc2} which, however, is not an exhaustive list.

\section{New localization principle}\label{new}
We discuss the new localization principle 
and give the first example of a localization deformation that
is obtained using it.

We define that 
\begin{equation}
(S(\lambda), \chi(\lambda), \psi)
\end{equation}
is a triple of the new localization principle (or triple)
if 
\begin{enumerate}
\item $S(\lambda)=H(\lambda)+\omega(\lambda)$ is a one-parameter family of
Hamiltonian structures with the exception that the symplectic
form can be degenerate except when $\lambda$ is zero, 
\item
$\chi(\lambda)$ is a one-parameter family of vector fields
satisfying \begin{equation} d_{\chi(\lambda)}S(\lambda)=0 \end{equation}
and 
\item$\psi$ is such a one-form that 
\begin{equation}
\frac{dS}{d\lambda}=d_{\chi(\lambda)}\psi.\label{localizationprinciple}
\end{equation}
\end{enumerate}
\begin{theorem} If $(S(\lambda), \chi(\lambda), \psi)$ is a triple
then the partition function
\begin{equation}
Z_{\lambda}=\int d\phi^{\mu}\:d\psi^{\mu}e^{S(\lambda)}\label{newpart}
\end{equation}
is independent of $\lambda$.
\end{theorem}
{\it Proof:} If one makes an infinitesimal change of variables in the direction of the supervector field $V=\psi d_{\chi(\lambda)}$,
the integrand is invariant and 
the Jacobian is $1+\epsilon \,Sdiv\:V=1+\epsilon \,d_{\chi(\lambda)}\psi$
which proves that $Z(\lambda+\epsilon)=Z(\lambda).\Box$

We make few remarks on the theorem that is the new localization principle.

{\it Remark 1:} The definition of the principle is complicated
due to the fact that one cannot assume that $\omega(\lambda)$
is nondegenerate for all $\lambda.$ If one could neglect the complication,
taking any point $H+\omega$ in the
space of Hamiltonian structures and any one-form $\psi$ on
the phase space
would give a one-parameter flow of Hamiltonian structures,
because then $\chi(\lambda)$ is uniquely determined as 
\begin{equation}
\chi(\lambda)=\omega(\lambda)^{-1}dH(\lambda).
\end{equation}

{\it Remark 2:} There are some interesting special cases of the localization
principle. If the one-form is exact $\psi=dF$ it reduces
to a canonical transformation generated by the function $F$.
On the other hand if $L_{\chi}\psi=0$ and $\chi(\lambda)=\chi$
it is the BRST localization principle.

{\it Remark 3:} The one-form $\psi$ is in many examples of the 
principle associated with
a phase space metric $g$ that is contracted
with some vector field, for example  $\psi=i_{\chi(0)} g$.

{\it Remark 4:} The loop space version of the principle is obtained by
thinking that the phase space is the infinite dimensional
loop space.

{\it Remark 5:} If $\omega(\lambda)=\omega(0)+\lambda d\psi$ is nondegenerate for all $\lambda$
the localization principle further simplifies.
We define another vector field
\begin{equation}
u(\lambda)=\omega(\lambda)^{-1}\psi
\end{equation}
and write the flow equation (\ref{localizationprinciple})
in the alternative form
\begin{equation}
\frac{dS}{d\lambda}=-L_{u(\lambda)}S\label{alternative},
\end{equation}
which is a diffeomorphism of the phase space. It
is then possible to transport any tensor  $T$
along the flow (\ref{alternative})
according to
the equation
\begin{equation}
\frac{dT}{d\lambda}=-L_{u(\lambda)}T.\label{alternative2}
\end{equation} 
One should not think that the invertibility
of $\omega(\lambda)$ is generic, for example the Duistermaat-Heckman theorem on a compact phase space\cite{own} gives a counterexample, which can be deduced\footnote{We thank A. Losev for this observation.} as follows: On a compact phase
space the Hamiltonian function has a maximum and minimum which cannot change
under any diffeomorphism flow. But in the Duistermaat-Heckman
case the maximum of the Hamiltonian
function goes to infinity under the localization flow because $H(\lambda)=H(0)+\lambda g(\chi(0),\chi(0))$, where $g$ is 
a phase space metric. Thus the flow cannot be a diffeomorphism
and therefore $\omega(\lambda)$ must be degenerate for some $\lambda$.

Finally, we give the first example of the localization 
principle. Suppose that the phase space admits a metric $g$
such that 
\begin{equation}
\nabla_{\chi}{\chi}=0.\label{geodesiccond}
\end{equation}
Then $(\frac{1}{2}g((1+\lambda)\chi,(1+\lambda)\chi)+d(i_{(1+\lambda)\chi}g), 
 (1+\lambda)\chi,i_{\chi}g)$ is a triple of the new localization principle
provided that the two-form $d(i_{\chi}g)$ is nondegenerate.
For example the geodesic motion on homogeneous manifolds has this structure, there
is actually a one-parameter family of metrics (\ref{phasespacemetric}) satisfying the
condition (\ref{geodesiccond}). As a triple it is a very special case because
$\lambda$ can be actually any function that satisfies $L_\chi\lambda=0$
and the partition function is still independent of $\lambda$.
The conditions of a triple can be proved using the identity\cite{own}
$$
\nabla_{\chi}\chi=0\Leftrightarrow d_{\chi}(\frac{1}{2}g(\chi,\chi)+
d(i_{\chi}g))=0.
$$

\section{Non-abelian localization principle}\label{non}
We review the non-abelian localization principle
making some additions to the original discussion in Ref.
\cite{2qc}. In the end of the section it is interpreted using
the new localization principle.

The partition function for two dimensional gauge theories is
\begin{equation}
Z=\int [ dA_{\mu}^{B} ] e^{-\frac{1}{e}\int tr\:F_{\mu\nu}F^{\mu\nu}}.
\label{2stars}
\end{equation}
The configuration space ${\cal A}$ of gauge potentials that is integrated over 
has a symplectic form 
\begin{equation}
\omega=\int dx^{\mu}\wedge dx^{\nu}\delta A_{\mu}^B\wedge\delta A_{\nu}^B
=\frac{1}{2}\int\sqrt{g}d^2x\:\sqrt{g}\epsilon^{\mu\nu}\delta A_{\mu}^B\wedge
\delta A_{\nu}^B,\label{1stars}
\end{equation}
a metric
\begin{equation}
g=\int\sqrt{g}d^2x \:g^{\mu\nu}\delta A_{\mu}^B\otimes\delta A_{\nu}^B
\end{equation}
and an almost complex
structure
\begin{equation}
J=\omega^{-1}g,\:\:\: J^2=-{\bf 1}.
\end{equation}
The partition function (\ref{2stars}) is 
actually an integration over the Liouville measure of the symplectic form (\ref{1stars}).
In addition the ``Hamiltonian'' in the partition function (\ref{2stars}) is the quadratic Casimir of the group of gauge transformations:
The gauge transformations act symplectically and the momentum map is
$\mu:{\cal A}\rightarrow \mbox {g}^*$ (g is the Lie algebra
of gauge transformations)
\begin{equation}
\mu(A)=F=dA+A\wedge A.
\end{equation}
In other words
\begin{equation}
\mu_{\epsilon}=\int F_{\mu\nu}^B\epsilon^Bdx^{\mu}\wedge dx^{\nu}
=\int\sqrt{g}d^2x\:(\sqrt{g}\epsilon^{\mu\nu}F^B_{\mu\nu})\epsilon^B
\end{equation}
generates the gauge transformation
\begin{equation}
\{ \mu_{\epsilon},A_{\mu}^B\}=D_{\mu}\epsilon^B.
\end{equation}
The associated Hamiltonian vector fields are 
\begin{equation}
v_{\epsilon}=\omega^{-1}d\mu_{\epsilon}=\int\sqrt{g}d^2x\:D_{\mu}^B\epsilon
\frac{\delta}{\delta A_{\mu}^B}
\end{equation}
which generate isometries of the metric
\begin{equation}
L_{v_{\epsilon}}g=0.
\end{equation}
Using the Hamiltonian generators of the Lie algebra of gauge transformations 
\begin{equation}
\mu^{Bx}=\mu_{\epsilon=\delta^{BC}\delta(x-y)}=\sqrt{g}\epsilon^{\mu\nu}
F_{\mu\nu}^B
\end{equation}
one can write the Hamiltonian in (\ref{2stars}) as the quadratic Casimir 
\begin{equation}
S=\int\sqrt{g}d^2x\:\mu^{Bx}\mu^{Bx}=\int\sqrt{g}d^2x\:F^{B}_{\mu\nu}
F^{\mu\nu}_B.
\end{equation}

The discussion above is analogous to the
following finite dimensional situation:
We assume that $C=\sum_i\mu_i^2$ is the quadratic Casimir
of a group that acts symplectically on the phase space 
and that there is a metric $g$ which is invariant.
The partition function\footnote{See Ref. \cite{2qc} for
the correct reinsertion of the imaginary unit.} is
\begin{eqnarray}
Z&=&\int \omega^n\:e^{-\sum\mu_i^2}=\int dx^{\mu}dc^{\mu}e^{-\sum\mu_i^2+\omega}\\
&=&\int d\phi_i\: e^{\frac{1}{4}\sum\phi_i^2}\int dx^{\mu}dc^{\mu}e^{\sum\phi_i\mu_i+\omega},
\end{eqnarray}
where $\sum\phi^i\mu_i +\omega$ is closed by the equivariant derivative
\begin{equation}
d+i_{\phi^iv_i}
\end{equation}
and $v_i$ is the Hamiltonian vector field of $\mu_i$.
We choose a one-form $\psi$ on the phase space that is invariant under
the action of the group,
\begin{equation}
L_{v_i}\psi=0
\end{equation}
for all $v_i$. One can use the BRST localization principle to prove that
\begin{eqnarray}
Z&=&\int d\phi_i\: e^{\frac{1}{4}\sum\phi_i^2}\int dx^{\mu}dc^{\mu}e^{\sum\phi_i\mu_i+\omega+\lambda(d+i_{\phi^iv_i})\psi}\\
&=&\int dx^{\mu}dc^{\mu}\:e^{-\sum_i(\mu_i+\lambda i_{v_i}\psi)^2+\omega
+\lambda d\psi}\label{4stars}
\end{eqnarray}
is independent of $\lambda.$ The limit $\lambda\rightarrow\infty$ yields a localization
on 
\begin{equation}
i_{v_i}\psi=0.\label{3stars}
\end{equation}
We choose more specifically $\psi=i_{\chi}g$ ($\chi=\{\sum\mu_i^2,\:\}=2\sum
\mu_iv_i$), which 
is equivalent to the one-form in Ref. \cite{2qc}, and get a localization
on the critical points of $\sum\mu_i^2$, which is
the non-abelian localization principle.

The equation (\ref{4stars}) can be interpreted as a consequence of the new localization
principle because $(-\sum_i(\mu_i+\lambda i_{v_i}\psi)^2+\omega
+\lambda d\psi,2\sum(\mu_i+\lambda i_{v_i}\psi)v_i, i_{\chi}g)$
is a triple. The conditions of a triple can be proved using the identity $(d+i_{v_i})(\mu_i+\omega+\lambda d_{v_i}\psi)=0.$ 
We remind that, neglecting possible complications due
to the degeneracy of $\omega(\lambda),$ one deforms 
the action $S=-\sum_i\mu_i^2+\omega$
using the flow (\ref{localizationprinciple}) where the one-form is
$\psi=i_{\chi}g$.

\section{Conformal Duistermaat-Heckman formula}
\label{con:sec}
We derive the conformal Duistermaat-Heckman formula
using the new localization principle. We have to
impose, however, a restriction that is not present in the original approach
in Ref. \cite{con}.

In the following it is assumed that the phase space admits a metric $g$ such that the Hamiltonian
vector field $\chi$ is a conformal Killing vector 
\begin{equation} 
L_{\chi}g=\Lambda g.
\end{equation}
$\Lambda$ is a function on the phase space.
In addition it is assumed that 
\begin{enumerate}
\item
there is a one-parameter family of vector fields 
$\chi(\lambda)$ such that
\begin{equation}
d_{\chi(\lambda)}(H+\omega+\lambda d_{\chi}(i_{\chi}g))=0\label{extra}
\end{equation}
and 
\item
\begin{equation}
g(\chi(\lambda),\chi)=g(\chi,\chi) \:\:\mbox{at the points $p$ where $\Lambda$ vanishes,}\label{asdfg}
\end{equation} 
which follows if  $\chi(\lambda)=\chi$ at the points p or if 
$\omega(\lambda)=\omega+\lambda d(i_{\chi}g)$
is nondegenerate for all $\lambda.$
\end{enumerate}

\begin{lem}
$(H+\omega+\lambda d_{\chi}(i_{\chi}g), \chi(\lambda), i_{\chi}g)$
is a triple of the new localization principle.
\end{lem}
{\it Proof:} We prove only the condition (\ref{localizationprinciple}) of a triple because the other two conditions are evident.
If $\lambda$ is zero (\ref{localizationprinciple})
is trivially true and we can assume that it is non-zero.
Conformal Killing vector satisfies the equation
$$dg(\chi,\chi)=\Lambda i_{\chi}g-i_{\chi}d(i_\chi g)$$ 
and using the condition (\ref{extra}) we get the identity
$$i_{\chi(\lambda)}\omega(\lambda)=-d(H+\lambda g(\chi,\chi))=i_{\chi}\omega
(\lambda)-\lambda\Lambda i_{\chi}g.$$
If $\Lambda\neq 0$ we get that
$$i_{\chi(\lambda)}i_{\chi}g=\frac{1}{\lambda\Lambda}i_{\chi(\lambda)}i_{\chi}
\omega(\lambda)
=\frac{1}{\lambda\Lambda}i_{\chi}dH(\lambda)=g(\chi,\chi),$$
which is enough to prove (\ref{localizationprinciple}).
If $\Lambda$ vanishes then (\ref{localizationprinciple})
follows directly from the restriction (\ref{asdfg}).$\Box$

Because of the new localization principle the partition function
\begin{equation}
Z_{\lambda}=\int d\phi^{\mu}d\psi^{\mu} e^{S(\lambda)}
\end{equation}
is independent of $\lambda,$ and in the limit $\lambda \rightarrow -\infty$
the action produces a delta function $\delta(\chi)$
which localizes the integral to the critical points of the Hamiltonian $H.$
Calculation of the integral gives the Duistermaat-Heckman formula
\begin{equation}
Z=\sum_{dH=0} \frac{\sqrt{\det
\omega_{\mu\nu}}}{\sqrt{\det\frac{\partial^2H}{\partial
\phi^{\mu}\partial \phi^{\nu}}}}e^{-\beta H}.\label{dh}
\end{equation}
We have used the fact that $\Lambda=0$ at the critical
points which follows from the formula \cite{con}
$$\Lambda=\frac{1}{2D}\nabla_{\lambda}\chi^{\lambda}=\frac{1}{2D}(\nabla_{\lambda}\omega^{\lambda\rho})\partial_{\rho}H.$$
The details of the calculation can be found in Ref. \cite{own}.

\section{Localization deformation of the Laplacian
on homogeneous spaces}\label{sec:lap}

We derive a localization deformation (\ref{eq50half}) of the Laplacian
(\ref{Jpartitionfunction}), (\ref{action}) on homogeneous spaces using the new localization
principle. It is demonstrated that the deformation does not give the desired
localization to the geodesics. However, it is worth presenting
as a non-trivial solvable example of the triple, or more accurately
combination of two of them. We mention also
another flow that might localize, but it seems to be non-solvable 
in closed form and it is not clear if the conditions of a triple are
satisfied (we conjecture that they are satisfied). See Appendix \ref{sec:geo}
and \ref{sec:pha} for notations.

We consider the partition function 
\begin{equation}
Z=\mbox{tr}\: e^{-\beta(\frac{1}{2} \triangle_0-J^i v_i)}\label{Jpartitionfunction}
\end{equation}
where $\beta$, $J^i$ are constants and $\triangle_0$ is the 
0-form part of the Laplacian $\triangle=dd^*+d^*d$ on homogeneous
manifolds. $e^{\beta J^iv_i}$ is the translation operator in the
direction of the isometry $J^iv_i.$ The partition
function (\ref{Jpartitionfunction}) coincides with the heat kernel that
is integrated over the manifold:
\begin{eqnarray}
Z&=&\int \sqrt{g}dx<x|e^{-\beta(\frac{1}{2}\triangle_0-J^iv_i)}|x>=\int\sqrt{g}dx
<x|e^{-\frac{1}{2}\beta\triangle_0}|e^{\beta J^iv_i}x>\\
&=&\int \sqrt{g}dx\:k_{\beta}(x,e^{\beta J^iv_i}x).
\end{eqnarray}

The path integral presentation of (\ref{Jpartitionfunction})
is
\begin{equation}
Z=\int[dx^{\mu}\,dp_{\mu}\,d\psi^{\mu}\,d\bar{\psi}_{\mu}]
\exp \int_0^{\beta}I_i\omega^i(\dot{x})-\frac{1}{2}K^{ij}I_iI_j
+J^iI_i+d(I_i\omega^i) \label{action}
\end{equation}
where $\frac{1}{2}K^{ij}I_iI_j=\frac{1}{2}g^{\mu\nu}p_{\mu}p_{\nu}$
is the Hamiltonian of the geodesic motion and the Grassmann variables are
associated with the one-forms, $\psi^{\mu}\sim dx^{\mu}$,
 $\bar{\psi}_{\mu}\sim dp_{\mu}.$
In addition there is DeWitt's term\cite{dew}, we assume that it
is proportional to the scalar curvature\footnote{There has been some
controversy about this term, see Ref. \cite{dew} for a recent
discussion and the speculation
in the end of Sec. \ref{conf}.}. We have neglected
it because homogeneous manifolds are of constant scalar curvature
and it yields only a shift in the energy levels.

The straightforward way to derive a
localization deformation of the 
action in equation (\ref{action}) would be to choose the loop space one-form
$$\psi=i_{\chi_S}g$$
where $g$ is the phase space metric (\ref{phasespacemetric}) (lifted\cite{wkb}
on the loop space) and $\chi_S$
is the loop space Hamiltonian vector field\cite{wkb},
$$\chi_S=\dot{\phi}-\chi+J^iv_i^H,\:\chi=\frac{1}{2}K^{ij}I_iv_j^H,$$
and apply the flow equation (\ref{localizationprinciple}) neglecting
the degeneracy problem. One can think (hopefully) that the inverse of the symplectic
form becomes singular at some points of the phase space but $\chi_S(\lambda)=\omega(\lambda)^{-1}dH(\lambda)$ stays
finite, we conjecture that this is the case.
However, we have not been able to solve the flow equation.
A power series solution experiment seems to give an infinity of
different terms which probably indicates that the flow is not solvable in closed
form. However, this flow might localize to the geodesics if one could
extract its asymptotic behaviour somehow.

The following refined approach gives a localization deformation that
is solvable and does not suffer from the degeneracy problem of the symplectic form, the latter is argued afterwards for reasons of pedagogy. 
We use
two one-forms ($g_1$ and $g_2$ are the two components of the metric
(\ref{phasespacemetric}))
$$\psi_i=i_{\chi_S}g_i,\:\:i=1,2$$
both of which give a flow that is exactly solvable and localizes half
of the degrees of freedom. In addition $\psi_2$ satisfies
\begin{equation}
L_{\chi_S}\psi_2=0\label{liederivativecondition}
\end{equation}
so that the flow in the direction of it is just a BRST flow
$$S\longrightarrow S+\alpha d_{\chi_S}\psi_2.$$

We combine the flows as follows: \\
 
\begin{figure}[h]
\centering
\setlength{\unitlength}{0.00066667in}
\begingroup\makeatletter\ifx\SetFigFont\undefined
\def\x#1#2#3#4#5#6#7\relax{\def\x{#1#2#3#4#5#6}}%
\expandafter\x\fmtname xxxxxx\relax \def\y{splain}%
\ifx\x\y   
\gdef\SetFigFont#1#2#3{%
  \ifnum #1<17\tiny\else \ifnum #1<20\small\else
  \ifnum #1<24\normalsize\else \ifnum #1<29\large\else
  \ifnum #1<34\Large\else \ifnum #1<41\LARGE\else
     \huge\fi\fi\fi\fi\fi\fi
  \csname #3\endcsname}%
\else
\gdef\SetFigFont#1#2#3{\begingroup
  \count@#1\relax \ifnum 25<\count@\count@25\fi
  \def\x{\endgroup\@setsize\SetFigFont{#2pt}}%
  \expandafter\x
    \csname \romannumeral\the\count@ pt\expandafter\endcsname
    \csname @\romannumeral\the\count@ pt\endcsname
  \csname #3\endcsname}%
\fi
\fi\endgroup
{\renewcommand{\dashlinestretch}{30}
\begin{picture}(4133,2833)(0,-10) 
\drawline(2846.000,2409.000)(2755.159,2439.379)(2663.682,2467.785)
	(2571.612,2494.206)(2478.992,2518.629)(2385.866,2541.043)
	(2292.276,2561.437)(2198.267,2579.802)(2103.883,2596.129)
	(2009.168,2610.410)(1914.166,2622.640)(1818.921,2632.811)
	(1723.479,2640.920)(1627.884,2646.962)(1532.180,2650.935)
	(1436.413,2652.837)(1340.628,2652.666)(1244.868,2650.424)
	(1149.179,2646.111)(1053.606,2639.729)(958.193,2631.281)
	(862.986,2620.771)(768.028,2608.204)(673.364,2593.586)
	(579.038,2576.923)(485.095,2558.224)(391.579,2537.497)
	(298.533,2514.752)(206.000,2490.000)
\drawline(2722.792,2419.940)(2846.000,2409.000)(2742.435,2476.633)
\put(1303,2734){\makebox(0,0)[lb]{\smash{{{\SetFigFont{9}{10.8}{rm}$\lambda$}}}}}
\put(1303,2450){\makebox(0,0)[lb]{\smash{{{\SetFigFont{9}{10.8}{rm}$\psi_1$}}}}}
\put(2927,2368){\makebox(0,0)[lb]{\smash{{{\SetFigFont{9}{10.8}{rm}$\psi_2(\lambda)$}}}}}
\put(0,2531){\makebox(0,0)[lb]{\smash{{{\SetFigFont{9}{10.8}{rm}$\psi_2$}}}}}
\put(0,2330){\makebox(0,0)[lb]{\smash{{{\SetFigFont{9}{10.8}{rm}$S$}}}}}
\put(2927,2180){\makebox(0,0)[lb]{\smash{{{\SetFigFont{9}{10.8}{rm}$S(\lambda)$}}}}}
\drawline(3138,2110)(3138,207)
\drawline(3108.000,327.000)(3138.000,207.000)(3168.000,327.000)
\put(2598,1106){\makebox(0,0)[lb]{\smash{{{\SetFigFont{10}{12.0}{rm}$\psi_2(\lambda)$}}}}}
\put(3228,1106){\makebox(0,0)[lb]{\smash{{{\SetFigFont{10}{12.0}{rm}$\alpha$}}}}}
\put(2841,30){\makebox(0,0)[lb]{\smash{{{\SetFigFont{10}{12.0}{rm}$S(\alpha,\lambda)$}}}}}
\end{picture}
}

\caption{Localization principle}\label{picture}
\end{figure}
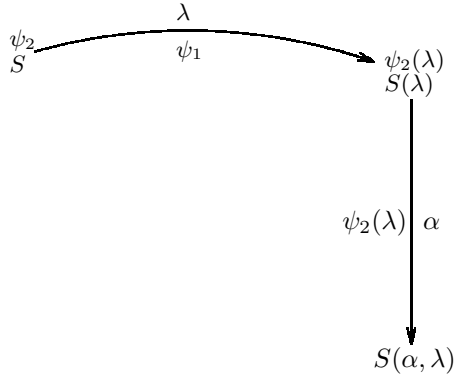

First, the
action and the one-form $\psi_2$ are evolved by the flow of the
one-form $\psi_1,$ the localization parameter of the flow is $\lambda.$
Secondly, the resulting action is evolved by the
flow of the transported one-form $\psi_2(\lambda)$ and
the localization parameter is $\alpha.$
The principle is described
graphically in Fig.~\ref{picture}, by the arrows there is the localization
parameter and the one-form which generates the flow.
It is important to note that the first flow turns out to be such that $\omega(\lambda)$ is nondegenerate
for all $\lambda.$ It can be seen using the formulas in Appendix~\ref{sec:geo} and~\ref{sec:pha}. Thus, one can transport the one-form $\psi_2$
along it according to the equation (\ref{alternative2}). In addition the flow preserves the Lie derivative condition (\ref{liederivativecondition})
in the form $$L_{\chi_S(\lambda)}\psi_2(\lambda)=0.$$ Consequently, the flow along the transported one-form $\psi_2(\lambda)$ is again a BRST flow.
The conditions of a triple are trivially satisfied because both the flows are familiar special cases: a diffeomorphism
and a BRST flow. In the former the vector field $\chi_S(\lambda)$ is 
obtained from  $\chi_S(0)$ by letting the diffeomorphism flow it
according to the equation (\ref{alternative2}).
In the latter $\chi_S(\alpha,\lambda)=\chi_S(\lambda).$

We get the total two-parameter localization deformation
\begin{eqnarray}
S(\alpha,\lambda)&=&S-S_B(\alpha,\lambda)-S_F(\alpha,\lambda)
\label{locaction}\\
S_B&=&(\lambda+\frac{1}{2}\lambda^2)K_{ij}\tilde{\chi}_x^i
\tilde{\chi}_x^j+\alpha K^{ij}\partial_t^JI_i(\lambda)\partial_t^JI_j(\lambda)\label{quadratic}\\
S_F&=&\lambda d(K_{ij}\tilde{\chi}_x^i\omega^j)+\alpha K^{ij}d(\partial_t^J I_i(
\lambda)dI_j(\lambda)),
\end{eqnarray}
where 
\begin{eqnarray}
\tilde{\chi}_x^i&=&\omega^i(\dot{x})-K^{ij}I_j+g^i_jJ^j\\
I_i(\lambda)&=&I_i-\lambda K_{ij}\tilde{\chi}_x^j\\
\partial_t^J&=&\delta^j_i\partial_t+J^kC^j_{ki}.
\end{eqnarray}

The partition function
\begin{equation}
Z(\alpha,\lambda)=\int[dx\,dp\,d\psi\,d\bar{\psi}]e^{S(\alpha,\lambda)}\label{eq50half}
\end{equation} is independent of the parameters $\alpha$ and $\lambda$ and coincides
with (\ref{action}) when they vanish. The limit that
the localization parameters are put
to infinity localizes
only half of the degrees of freedom, namely 
\begin{equation}
\tilde{\chi}_x^i=0\label{locahalf},
\end{equation} leaving an infinite dimensional integral. The full
localization on the equations of motion (the geodesics) would be 
$$\partial_t^JI_i=0,\:\:\tilde{\chi}_x^i=0.$$

The fact that the localization deformation
localizes only half of the degrees of freedom can be proved as follows:
As $\lambda$ and $\alpha$ are large there is
exponential damping in the path integral due to the bosonic
part (\ref{quadratic}), unless
\begin{equation}
\tilde{\chi}_x^i\sim\frac{1}{\lambda}f^i,\:\:\:
\partial_t^JI_i\sim\partial_t^JK_{ij}f^j+\frac{1}{\sqrt{\alpha}}g_i\label{estimates}
\end{equation}
where $f^i$ and $g^i$ are finite.
The fermionic part does not change the situation because
the zeta function scaling of the fermionic determinant
shows that it can only give a polynomial dependence.
We see from the equations (\ref{estimates}) that we get
localization only on (\ref{locahalf}).
However, it might be that taking first $\lambda$ to infinity
and then $\alpha$ would localize, or although there is
no exponential damping, there might be a rational delta function
$\delta(x)\sim\frac{\alpha}{1+(\alpha x)^2}.$ 
Neither happens, as can be proved in the case $J=0$ by integrating the fermions and
expanding the bosonic action
around $\tilde{\chi}_x^i=0$ 
using the coordinate system 
$x^\mu,\,\phi_S^{\mu}=v_i^{\mu}\tilde{\chi}_x^i.$
One sees that the residual $\alpha$ and $\lambda$ dependence cannot 
localize further. In addition, if there would be localization in the
case that $J\neq0,$ there would also be localization in the limit that
$J$ vanishes.

\section{Non-abelian derivative expansion localization}\label{ano}
We localize the heat kernel on compact Lie groups
using a derivative expansion localization. In the
end of the section it is commented how it might be possible to obtain
new localization formulas using the principle. Many of the equations in this section are formal,
they involve distributions or possess
singularities that cancel. We do not address the difficult mathematical
problem of how to treat them rigorously.

We begin by studying the shifted heat kernel
\begin{equation}
<x|e^{-\beta (\frac{1}{2}K^{ij}v_iv_j-J^iv_i)}|y>
\end{equation}
 on homogeneous manifolds $M=G/H.$ If one puts the points $x,\,y\in M$ equal and integrates
over $x$ one gets the partition function (\ref{Jpartitionfunction}).

\begin{theorem}\label{theorem2}Provided that the series (\ref{series1}) converges in the sense of distribution theory\cite{rud}
the following formal identity holds:
\begin{equation}
<x|e^{-\beta (\frac{1}{2}K^{ij}v_iv_j-J^iv_i)}|y>=\left.e^{- \frac{1}{2}\beta K^{ij}u_i u_j}
<x|e^{\varphi^iv_i}|y>\right|_{\varphi=\beta J}\label{identityx}
\end{equation}
where\footnote{$(C_i)^j_k=C^j_{ik}$ are the structure constants, see
Appendix \ref{sec:geo}.}
\begin{equation}
u_i=\left(\frac{-\varphi^kC_k}{e^{-\varphi^kC_k}-1}\right)^j_i\frac{\partial}
{\partial \varphi^j}.\label{ui}
\end{equation}
\end{theorem}
{\it Proof:} We expand the exponential function on the left hand side as
\begin{equation}
<x|e^{-\beta (\frac{1}{2}K^{ij}v_iv_j-J^iv_i)}|y>=
\sum_{n=0}^{\infty}<x|e^{\varphi^iv_i}(-\frac{1}{2}\beta K^{ij}v_iv_j)^n|y>/n!\label{series1}\end{equation}
and by definition the right hand side is
\begin{equation}
e^{- \frac{1}{2}\beta K^{ij}u_i u_j}<x|e^{\varphi^iv_i}|y>=
\sum_{n=0}^{\infty} (- \frac{1}{2}\beta K^{ij}u_i u_j)^n/n!\,<x|e^{\varphi^iv_i}|y>.\label{series2}
\end{equation}
The terms in the series on the right hand side are actually distributions:
$n=0$ term is
\begin{equation}
<x|e^{\varphi^iv_i}|y>=\delta(x-e^{\varphi^i v_i}y),
\end{equation}
and higher terms are obtained by differentiating it (it may
happen that the terms diverge because of
the singularities of $u_i$ but then the series does not converge).
Moreover, the terms on both sides are equal in each order (which suffices to prove the theorem because the series converge) because of the formal identity
\begin{equation}\label{qwer}u_i\,e^{\varphi^kv_k}=e^{\varphi^kv_k}v_i\end{equation}
that can be proved as follows: Using Duhamel's formula\cite{ber}
\begin{equation}
\frac{\partial}{\partial\varphi^i}e^{\varphi^kv_k}=e^{\varphi^kv_k}\omega(\varphi)^j_iv_j
\end{equation}
where
\begin{equation}
\omega(\varphi)^j_id\varphi^i =\left(
\frac{e^{-\varphi^kC_k}-1}{-\varphi^kC_k}\right)^j_id\varphi^i
\end{equation}
can be understood
as the left-invariant one-form of the Lie group $G$ in exponential coordinates.
It is, however, continued on the whole Lie algebra including the points
where the coordinate system is singular.
The equation (\ref{qwer}) follows inverting it, which is possible
except at the singular points of the exponential coordinate
system. However, formally the equation
(\ref{qwer}) holds also at these points, then
$u_i$ is singular but the singularity cancels in the whole expression (\ref{qwer}).$\Box$
 
One can turn the derivative expansion in Theorem \ref{theorem2}
formally into an integral using the kernel
\begin{equation}
K_{\beta}(\varphi',\varphi)=<\varphi'|e^{-\frac{1}{2}\beta K^{ij}u_iu_j}|\varphi>
\end{equation}
as follows:
\begin{equation}e^{- \frac{1}{2}\beta K^{ij}u_i u_j}
<x|e^{\varphi^iv_i}|y>=\int\sqrt{g'(\varphi')}\,d\varphi' K_{\beta}(\varphi,\varphi')<x|e^{\varphi'^iv_i}|y>\label{integralloc}
\end{equation}
where $g'$ is the degenerate
metric 
\begin{equation}
g'(\varphi)_{kl}=K_{ij}\omega(\varphi)^i_k\omega(\varphi)^j_l
\end{equation}
on the Lie algebra. 

In addition one can formally calculate the kernel as follows:
$$K_{\beta}(\varphi',\varphi)=<0|e^{-\frac{1}{2}\beta K
^{ij}u_iu_j}|\xi(\varphi',\varphi)>=K_{\beta}(0,\xi)$$
where $\xi$ is defined as
\begin{equation}
e^{-\varphi'^kv_k}e^{\varphi^kv_k}=e^{\xi(\varphi',\varphi)^kv_k}.
\end{equation}
An explicit expression for $\xi$ can be obtained using the Campbell-Baker-Hausdorff theorem\cite{sat} in a neighbourhood that $\varphi'$ and $\varphi$
are close to zero.
The kernel $K_{\beta}(0,\xi)$ has been calculated in Ref. \cite{mar},
\begin{equation}
K_{\beta}(0,\varphi)=
\frac{M}{(2\pi\beta)^{\frac{D}{2}}}\hat{A}(\varphi^iC_i)e^{-\frac{1}{2\beta}K_{ij}\varphi^i\varphi^j+\frac{D}{48}\beta}
\end{equation}
where $\hat{A}(X)=\prod_{x_k>0}\frac{x_k/2}{\sin x_k/2}$
($ix_k$ are the eigenvalues of the antisymmetric real matrix $X$) and $M$ is a normalization constant\cite{mar}\cite{pic}. 

We localize the Laplacian on Lie groups assuming that the condition
of Theorem \ref{theorem2} holds and one can transform the derivative
operator into an integral as in equation (\ref{integralloc}), which
is plausible because the singularities cancel.
We calculate
\begin{eqnarray}
k_{\beta}(\mbox{\bf 1},g)&=&<\mbox{\bf 1}|e^{-\frac{1}{2}\beta\triangle_0}|g>\\&=&
\int\sqrt{g'}\,d\varphi\,K_{\beta}(0,\varphi)\delta(\mbox{\bf 1}-ge^{\varphi^iT_i})=\sum_{\varphi\in L}K_{\beta}(\varphi),\label{heatkernelsemi}
\end{eqnarray}
where $\mbox{\bf 1},\,g\in G$ and we use the notation of matrix groups. $T_i$ are
the generators of the Lie algebra. One should notice that the delta function
is normalized with respect to the volume on the Lie group whereas the integral
is over $\varphi$, however, because one can associate $\varphi$ as the exponential coordinate we get agreement. The lattice\footnote{We have used the symmetry $K_{\beta}(\varphi)=K_{\beta}(-\varphi)$ to fix the sign convention for the lattice $L$.} $L$ is 
\begin{equation}
L=\{\varphi|g=e^{\varphi^iT_i}\}\label{L}.
\end{equation}
It is in one to one correspondence with the geodesics starting at {\bf 1} and ending at $g$ because one can associate with $\varphi\in L$ a geodesic $e^{\frac{t}{\beta}\varphi^iT_i}$, $t\in [ 0,\beta].$ 
The expression (\ref{heatkernelsemi}) coincides with the semiclassical approximation and it is studied in more detail in the next section.

Finally, we comment how it might be possible
to obtain new localization formulas using Theorem \ref{theorem2}.
Putting $y=x$ in Theorem \ref{theorem2} and integrating
over $x$ gives
\begin{equation}
\mbox{tr}\,e^{-\beta (\frac{1}{2}K^{ij}v_iv_
j-J^iv_i)}=\left.e^{- \frac{1}{2}\beta K^{ij}u_i u_j}
\mbox{tr}\,e^{\varphi^iv_i}\right|_{\varphi=\beta J}.\label{uiop}
\end{equation}
There is, however, a minor subtlety: the condition of Theorem \ref{theorem2}
does not necessary hold. In the Lie group case the terms in the series (\ref{series1}) are not in the space of distributions because the delta functions restrict the
vector field $u_i$ to its singular points, hence, the series
does not converge in the sense of distributions. The problem can be circumvented by replacing $y$ instead by $e^{\varphi_0^iv_i}x$ and continuating analytically $\varphi_0$ to zero (or modifying the formulas by $\varphi_0$).

The linear partition function 
\begin{equation}Z=\mbox{tr} \,e^{\varphi^kv_k}\label{J}\end{equation}
is in the path integral form($\varphi=\beta J$)
\begin{equation}
Z=\int[dx\,dp\,d\psi\,d\bar{\psi}]\exp\int_0^{\beta}I_i\omega^i(\dot{x})
+J^iI_i+d(I_i\omega^i).\label{123}
\end{equation}
It can be localized using 
the BRST localization principle because the Hamiltonian vector field of the Hamiltonian $J^iI_i$ generates
an isometry of the phase space metric (\ref{phasespacemetric}).
Consequently, we should get a localization for the partition function
(\ref{uiop}) that is a derivative expansion of
ordinary localization formulas. (This principle appeared first in the abelian
form in Ref. \cite{witloc}. There is also a path integral derivation
of it which lacks for the non-abelian case, see also Ref. \cite{tir2}.) 
However, one has to be careful because the linear partition function
(\ref{J}) is a distribution, therefore the
path integral (\ref{123}) must diverge (for example the divergent
integral  $\int_{-\infty}^{\infty}e^{ipx}dx$ can be associated with the delta function).

One can formally add a BRST exact term to the action in the path
integral (\ref{123})
with any of the following gauge fermions or their
linear combinations, multiplied
by the localization parameter $\lambda,$
\begin{eqnarray}
\psi_1&=&i_{J^iv_i^H}g\label{first}\\
\psi_2&=&i_{\dot{\phi}}g\label{second}\\
\psi_3&=&i_{\dot{\phi}-K^{ij}I_iv_j^H}g\label{third}
\end{eqnarray}
where $g$ is the phase space metric~(\ref{phasespacemetric}).

For example the gauge fermion (\ref{second}) should give
a localization formula that is an integral over the phase space\cite{tir} of an equivariant characteristic class\footnote{The Dirac genus should possibly be replaced by the Todd
class\cite{mauri} in Ref. \cite{tir}.}. The integral
must be divergent, but if one would be able to differentiate
the derivative expansion in the full formula (\ref{uiop}) one would perhaps get a convergent integral over
the phase space and a novel localization formula.

Moreover the gauge fermion $\psi_2-\psi_1$ should give
the loop space Duistermaat-Heckman formula\cite{wkb} or its degenerate
version, but it turns out to be singular because there are no critical points except
at fixed values of $\beta$. The localization
by the gauge fermion (\ref{second}) seems to be even more
singular, it should localize to the non-existing critical points of the Hamiltonian\cite{tir2} . The gauge fermion (\ref{third}) can possibly give a localization
only if one first sums the derivative expansion and then takes the limit,
which we have not been able to do.

\section{Configuration space localization}\label{conf}
We localize the Laplacian on a Lie group using a configuration
space path integral. 

On the space $\Omega G$ of based loops
(loops that start and end at the unit element {\bf 1} of the Lie group) there
is a natural symplectic form that is right-invariant\footnote{The left- and right-invariant forms have changed places
compared to Ref. \cite{pressley}.}
\begin{equation}
\omega=\frac{1}{2}K_{ij}\omega_R^i\wedge\partial_t\omega_R^j,\:\:d\omega=0.
\end{equation}
The notation is that $v_i, v_i^R$ are the left- and right-invariant
vector fields and the dual one-forms are
$$\omega^i=K^{ij}g(v_j),\:\: \omega^i_R=K^{ij}g(v_j^R)$$
where $g$ is the bi-invariant metric and the vector fields coincide at 
{\bf 1}, $\left.v_i\right|_{\mbox{\bf 1}}=\left.v_i^R\right|_{\mbox{\bf 1}}$.

The configuration space path integral is
\begin{equation}
Z=\int[dx^{\mu}\,d\psi^{\mu}]_{vbc}\exp\int_0^{\beta}\frac{1}{2}K_{ij}
(\omega^i(\dot{x})+J^i)(\omega^j(\dot{x})+J^j)+\frac{1}{2}K_{ij}\omega_R^i\wedge
\partial_t\omega_R^j \label{y}
\end{equation}
where $vbc$ means vanishing boundary conditions
for both bosons and fermions, $x^{\mu}(\beta)=
x^{\mu}(0)=0,\:\psi^{\mu}(\beta)=\psi^{\mu}(0)=0,$
and the coordinates $x^{\mu}$
are chosen so that $x^{\mu}=0$ corresponds
to the unit element {\bf 1} of the Lie group.

Integration of the fermionic part of the integral gives a Pfaffian
\begin{equation}
\sqrt{\det(\omega^i_{R\mu}K_{ij}\partial_t
\omega^j_{R\nu})}=\sqrt{\det g_{\mu\nu}}\,\sqrt{\det(\delta^i_j\partial_t)_{vbc}}\label{fermionicpart}
\end{equation}
where we have used the product rule for the determinant and the determinants
obey vanishing boundary conditions.
The equation (\ref{fermionicpart}) can be justified using the change of variables
\begin{equation}
\psi^{\mu}=v_i^{\mu}M^{-1}_{ij}\theta^j,\:\: K_{ij}=(M^TM)_{ij},
\end{equation} 
writing the measure as $[d\theta^j]_{vbc}=[d\theta^j]_{pbc}\delta
(\theta^j(0))$
and expanding $\theta^j$ in Fourier modes (pbc means periodic boundary 
conditions). The procedure also evaluates the determinant
$$\det(\delta^i_j\partial_t)_{vbc}=\mbox{det}'(\beta\partial_t)^D=1$$
in terms of the standard determinant\footnote{We have chosen a regularization that preserves the
``charge conjugation'' symmetry\cite{elitzur}\cite{mauri} $\mu\leftrightarrow -\mu.$}
\begin{equation}
\mbox{det}'(\partial_t+\mu)_{pbc}=\prod_{n\neq 0}\left(\mu+\frac{2\pi n i}{\beta}\right)=\beta\frac{\sinh \beta \mu/2}{\beta\mu/2}.\label{standarddet}
\end{equation}

The result, after the integration of fermions, is
$$
\int[\sqrt{g}dx^{\mu}]_{vbc}\,\exp\int\frac{1}{2}K_{ij}(\omega^i(\dot{x})+J^i)(\omega^j(\dot{x})+J^j)
$$
which using the change of variables 
\begin{equation}
x^{\mu}\longrightarrow e^{-tJ^iv_i}x^{\mu}\label{changevar}
\end{equation}
gives the equivalent path integral   
$$\int[\sqrt{g}\,dx]_{bc}\,\exp\frac{1}{2}\int^{\beta}_0K_{ij}
\omega^i(\dot{x})\omega^j(\dot{x})$$
with the boundary conditions (bc)
$$x^{\mu}(\beta)=\left.(e^{\beta J^iv_i}x^{\mu})\right|_{x=0}\sim e^{\beta J^iT_i}=g,\:\:
x^{\mu}(0)=0\sim\mbox{\bf 1}.$$
Thus, the configuration space path integral (\ref{y}) coincides with the heat kernel $k_{\beta}(\mbox{\bf 1},g).$

We proceed to localize the
integral (\ref{y}) using the Hamiltonian BRST localization principle.
The Hamiltonian vector field for the action\cite{pressley}
\begin{equation}
S=\int_0^{\beta}\frac{1}{2}K_{ij}
(\omega^i(\dot{x})+J^i)(\omega^j(\dot{x})+J^j)
\end{equation}
is 
\begin{equation}
\chi_S=\chi+J^iu_i
\end{equation}
where
\begin{eqnarray}
\chi&=&\dot{x}-\left.\omega^i(\dot{x})\right|_{t=0}v_i^R\\
u_i&=&v_i-v_i^R
\end{eqnarray}
and one can use the Hamiltonian BRST symmetry with the equivariant
derivative $d_{\chi_S}$.

We construct the gauge fermion $\psi$ starting with an invariant tensor
\begin{equation}
g'=\int_0^{\beta}K_{ij}d\omega^i(\dot{x})\otimes d\omega^j(\dot{x}),
\:\:L_{\chi_S}g'=0
\end{equation}
and contracting it with the loop space Hamiltonian vector field
\begin{equation}
\psi=i_{\chi_S}g'.
\end{equation}
This is analogous to the localization in Ref. \cite{wkb}.
Adding the gauge fermion gives the action
\begin{eqnarray}
S(\lambda)&=&S+\omega+\lambda d_{\chi_S}(i_{\chi_S}g')\\
&=&
S+\omega+\lambda K_{ij}\partial_t^J\omega^i(\dot{x})
\partial_t^J\omega^j(\dot{x})+\lambda K_{ij}d(\partial_t^J\omega^i(\dot{x})
d\omega^j(\dot{x}))\label{locaction2}
\end{eqnarray}
where  $\partial_t^J=\delta^i_j\partial_t
-J^kC^i_{kj}$. 

The path integral 
\begin{equation}
Z=\int[dx^{\mu}d\psi^{\mu}]_{vbc}e^{S(\lambda)}\label{locpathint}
\end{equation}
coincides with (\ref{y}) when
$\lambda$ vanishes and localizes in the limit $\lambda\rightarrow-\infty$ on $\partial_t^J\omega^i(\dot{x})=0$
which are the classical geodesics $\partial_t\omega^i(\dot{x})=0$  
before the change of variables in equation (\ref{changevar}). 

One can
write the geodesics starting at {\bf 1} and ending at $g=e^{\beta J^iT_i}$ (in matrix group notation)
as\cite{mar}\cite{pic} 
$$\gamma_{\nu}(t)=e^{\frac{t}{\beta}(\varphi^i+2\pi\nu^i)T_i},\:\:
\nu^i\in L$$
where $\varphi^i=\beta J^i$ and
$$L=\{\nu^i| [\nu^iT_i,\,\varphi^jT_j]=0,\:e^{2\pi\nu^iT_i}=
\mbox{\bf 1}\}$$
and we assume that $\varphi^iT_i$ is in the generic position that its annihilator is a conjugated Cartan subalgebra. One can also write
the lattice $L$ in terms of the roots of the algebra, see Refs. \cite{mar}
and \cite{pic}.
The solutions of the equation $\partial_t^J\omega^i(\dot{x})=0$ are
obtained from these by inverting the change of variables in equation
(\ref{changevar}) (which is
translated into matrix group notation as $x^{\mu}(t)\sim g(t)\rightarrow e^{-tJ^iv_i}x^{\mu}\sim g(t)e^{-tJ^iT_i}$)
 giving
\begin{equation}
\tilde{\gamma}(t)=e^{\frac{t}{\beta}2\pi\nu^iT_i}.\label{critical}
\end{equation}

Then we calculate the limit $\lambda\rightarrow-\infty$ 
in the path integral (\ref{locpathint}).

First we make the change of variables
$$[dx^{\mu}\,d\psi^{\mu}]_{vbc}\longrightarrow[dX^id\eta^i]$$
where $X^i=\omega^i(\dot{x})$ and $\eta^i=dX^i=\frac{\delta\omega^i(\dot{x})}{\delta x^{\mu}}
\psi^{\mu}$. It changes the vanishing boundary conditions into something more complicated. Fortunately it is
enough to study what happens near the critical points because of the localization and the fact that the first variation of the bosonic part in $S(\lambda)$ vanishes.
In first order the bosonic coordinates are related by the identity   
$$x^{\mu}=x^{\mu}_{cr}+\delta x^{\mu}, \:\delta\omega^i(\dot{x})=
\left.\frac{\delta\omega^i(\dot{x})}{\delta x^{\mu}}\right|_{x_{cr}}
\delta x^{\mu}=(\tilde{D}_t^{M-C})^i_j(\omega^j_{\mu}\delta x^{\mu})$$
where we have the Maurer-Cartan differential operator (\ref{mautilde}). 
The critical points are given in the equation (\ref{critical})
so that the Maurer-Cartan operator is $$\tilde{D}_t^{M-C}=e^{-\frac{t}{\beta}2\pi\nu^iC_i}\partial_te^{\frac{t}{\beta}2\pi\nu^iC_i}.$$
We get that the vanishing boundary condition results in the condition
\begin{equation}
\int_0^{\beta}e^{\frac{t}{\beta}2\pi\nu^iC_i}\delta X^i=0\label{bccondition}
\end{equation}
for the coordinate $X^i=X^i_{cr}+\delta X^i$ near the critical point.
The fermionic coordinates $\eta^i$ obey also the same
condition (\ref{bccondition}).

Integrating the fermions and putting the localization
parameter to infinity yields 
$$Z=\sum_{\nu}\frac{1}{\mbox{Pf}(\partial_t^J)}
e^{-\frac{1}{2\beta}K_{ij}(\varphi^i+2\pi\nu^i)(\varphi^j+2\pi\nu^j)}$$
where the Pfaffian obeys the boundary condition (\ref{bccondition}).
It is calculated conjugating
the Hilbert space by the operator $e^{-\frac{t}{\beta}
2\pi\nu^iv_i}$ which turns the boundary condition (\ref{bccondition})
into $\int_0^\beta\delta X^i=0$ and the operator $\partial_t^J$
into $\partial_t^{J+2\pi\nu}$.
Furthermore, the diagonalization of $(\varphi^i+2\pi\nu^i)C_i$ gives the eigenvalues $\mu_j$
\begin{equation}
\mu_j=
\left\{
\begin{array}{ll}
0;&j=1,\ldots,r\\
i<\varphi+2\pi\nu,\alpha_j>;&j=r+1,\ldots D
\end{array}
\right.
\end{equation}
where $r$ is the rank of the group, $\alpha_i$ are the roots (they are in
the annihilator of $\varphi^iT_i$ which is isomorphic to the Cartan subalgebra) and $<\:,\:>$ is the contraction by
the Killing tensor $K_{ij}$.
The Pfaffian reduces into a product of usual determinants (\ref{standarddet})
and the final result is
\begin{equation}
k_{\beta}(\mbox{\bf 1}, e^{\varphi^iT_i})
= \frac{M}{(2\pi\beta)^{\frac{D}{2}}}\sum_{\nu}
\prod_{\alpha>0}\frac{<\varphi+2\pi\nu,\alpha>/2}      
{\sin <\varphi+2\pi\nu,\alpha>/2}e^{-\frac{1}{2\beta}<\varphi+2\pi\nu,\,\varphi+2\pi\nu>}
\end{equation}
which however has to be corrected with DeWitt's term which gives
the extra factor $\exp\left
( -\beta\frac{D}{48}\right).$ It is consistent
with DeWitt's original\cite{dewori} proportionality constant $\frac{1}{12}$ in front of the scalar curvature $\frac{D}{4}.$ 
The result coincides with the expression (\ref{heatkernelsemi}) and has been calculated using different methods in Refs. \cite{mar} and \cite{pic}.
This is, however, disturbing because the natural value\cite{dew} of DeWitt's
constant should be $\frac{1}{8}$. We hope that further research solves this puzzle. We speculate
that both values may be correct, they just correspond to different path
integral measures, and that this trivial looking problem may
reveal deep insights in how to make the path integral rigorous and
what the quantization is all about.

\section{Conclusions, speculations and future prospects}\label{sec:conc}
We have developed two new localization methods: the new localization principle and the non-abelian
derivative expansion localization. We emphasize that both the principles
are universal and probably have many other applications which
are not yet known. For example
the non-abelian derivative expansion localization may apply to integrable
models: Many integrable models can be embedded in Poisson-Lie algebras\cite{rmatrix}.
The integrable hierarchy is the sequence of Casimirs with respect to the
r-deformed Poisson bracket. The linear generators generate the
coadjoint action and are therefore isometries of the Killing
metric on the algebra. However, the algebra in many cases
is non-compact and the metric that is pulled back on the phase space
may be indefinite. There may also be anomalous problems
associated with the quantization and the derivative expansion
in Theorem \ref{theorem2} may be singular, so that before concrete examples are worked out we cannot say if the principle eventually works or not.

Another direction of research is to generalize the configuration space localization on
Lie groups in section \ref{conf} to homogeneous or locally
homogeneous manifolds and try to understand Selberg's trace formulae.
However, it may be that the existence of Selberg's trace formulae is a separate phenomenon, see the remarks below. Yet another direction of research
would be to study if there would be an analogue of the configuration space localization for two dimensional sigma models, for example
the Laplacian on loop groups might be such.
In addition it and the derivative expansion localization in section \ref{ano} may have supersymmetric versions, the starting point would be 
the grand canonical partition function that was introduced in Ref. \cite{dew}.

Finally, we collect few remarks on Selberg's trace
formula on constant negative curvature Riemann surfaces which might
be helpful in a possible path integral localization.  

{\it Remark 6:} Selberg's trace formula is exact Gutzwiller's trace
formula\cite{gut2} which suggests that the path integral should perhaps be
enlarged by one degree of freedom
\begin{equation}
Z(\beta)=\mbox{tr}\,e^{-\beta\triangle_0}\rightarrow\int_0^{\infty}e^{-\beta E}Z(\beta).
\end{equation}

{\it Remark 7:} The Riemannian surfaces have a natural symplectic form,
the volume form $\sqrt{g}\epsilon_{\mu\nu}dx^{\mu}\wedge dx^{\nu},$ which
makes it possible to write the path integral
as follows:
\begin{equation}
Z(\beta)=\int [\sqrt{g}dx]\exp\int g_{\mu\nu}\dot{x}^{\mu}\dot{x}^{\nu}=\int
[dx^{\mu}\,d\psi^{\mu}]\exp\int g_{\mu\nu}\dot{x}^{\mu}\dot{x}^{\nu}+\sqrt{g}\epsilon_{\mu\nu}\psi^{\mu}\psi^{\nu}.
\end{equation}
Perhaps it should be taken to be the starting point for localization.

{\it Remark 8:} We have assumed in {\it Remark 7} that the path integral takes implicitly care of the fact that the Riemann surfaces are not simply connected.
It can be made explicit (it may even be that it is necessary) by integrating on the Poincar\'e upper half-plane and splitting the integral into a sum of path integrals with different boundary conditions. It is possible to transform the boundary conditions into periodic
using an analogous change of variables that was used in Sec. \ref{conf} in equation (\ref{changevar}), which allows one to use localization deformations. 
Furthermore, in order to replace the sum as an integral, one can perhaps make an analogous trick\cite{dit} as in the case of the material particle on $U(1)$:
\begin{eqnarray*}
\mbox{tr}\, e^{\beta\partial_x^2}&=&\sum_{n=-\infty}^{\infty}\int_0^Ldx(0)
\int[dx]_{x(\beta)=x(0)+nL}\mbox{e}^{\int\dot{x}^2}\\
&=&\int[dx]_{pbc}\exp\int\left(\dot{x}+\left[\frac{x(0)}{L}\right]L\right)^2,
\end{eqnarray*}
where $x(t)\in\mbox{\bf R}^1$ and $[x]$ is the greatest integer
smaller than $x$. $U(1)$ is associated with $\mbox{\bf R}^1 \:\mbox{mod}\: L.$

\section{Acknowledgements}
We thank A. Niemi for pointing out the reference \cite{con} and
the connection with Selberg's trace formula. We thank him also for the problem of finding a non-abelian
derivative expansion localization in connection
with integrable models. We thank O. Tirkkonen for pointing out the possibility of a configuration space localization.   
We thank K. Palo for giving his beautiful calculations
in the degenerate Duistermaat-Heckman localization. 
In addition we thank
A. Alekseev, Cheeqer, B. S. DeWitt, L. D. Faddeev, J. Kalkkinen, A. Losev, M. Miettinen, 
T. Nassar, A. Niemi, K. Palo, P. Pasanen, Rubtsov, R. J. Szabo and  
O. Tirkkonen for discussions. 

\appendix
\Section{Formulas on homogeneous spaces}\label{sec:geo}
We introduce formulas on homogeneous spaces that use a degenerate basis
for vector fields. In addition we define a Maurer-Cartan connection
on the tangent bundle.

A homogeneous space is, in this article, the quotient space $M=G/H$
($G$ and $H$ are Lie groups) with the metric that is inherited
from the unique bi-invariant metric on the Lie group $G$. Technically,
one inverts the metric on $G$ and pushes it on $M$ using the canonical projection\cite{coq}. We require in addition that $G$ is compact,
although the formulas are valid also in the non-compact case
with the exception that the metric tensors are not necessarily positive-definite.  

As the quotient is taken by the right-action we get the standard
left-action of $G$ on M which is generated by the vector fields
\begin{equation}
v_i,\:\: i=1,\dots,N\label{vecs}
\end{equation}
on the $D$ dimensional manifold $M$.
They are isometries of the metric and satisfy
the Lie algebra of $G$
\begin{equation}
[v_i,v_j]=C^k_{ij}v_k.\label{algebra}
\end{equation}
We define the Killing tensor
\begin{equation}
K_{ij}=-tr\:C_iC_j=-C^k_{il}C^l_{jk}
\end{equation}
that is positive definite because $G$ is compact. The inverse
of $K$ is $K^{ij}$ and it is used to raise indices $i,j,k\dots\in
\{1,\ldots,N\}$ as Greek indices $\alpha,\beta \dots\in\{1,\ldots,D\}$ 
are raised and lowered by the metric $g_{\mu\nu}$.
For example the tensor
$$C_{ijk}=K_{im}C^m_{jk}$$
is antisymmetric in all the three indices when the upper index is lowered. 

We define the 1-forms
\begin{equation}
\omega^i=K^{ij}g(v_j)
\end{equation}
that are dual to the vector fields (\ref{vecs}). If $x^{\mu}$ are coordinates on $M$ one can write the vector fields 
and their dual 1-forms in the coordinate basis as $v_i=v_i^{\mu}\partial_{\mu}$ and
$\omega^i=\omega^i_{\mu}dx^{\mu}$ which satisfy the relation
\begin{equation}
\omega^i_{\mu}v_i^{\nu}=\delta_{\mu}^{\nu}.
\end{equation}
The metric can be written as 
\begin{equation}
g=K_{ij}\omega^i\otimes\omega^j,\:\:\: g^{-1}=K^{ij}v_i\otimes v_j.
\end{equation}

We define a tensor $g^i_j$
\begin{equation}
g^i_j=\omega^i\cdot v_j
\end{equation}
that has the properties
\begin{equation}
g_{ij}=g_{ji},\: g^i_jg^j_k=g^i_k,\:g^i_j\omega^j=\omega^i,\: g^i_jv_i=v_j.
\end{equation}
Using it we can associate the Lie algebra
of the local isotropy group
$$H_p=\{g\in G|gp=p\},\:\: p\in M$$
with $$h_p=\{ X^iv_i|X^i\in \mbox{\bf R},\:g^i_j(p)X^j=0\},$$
and the orthogonal complement of it, that is isomorphic to the tangent space at $p$, with
$$m_p=\{X^iv_i|X^i\in\mbox{\bf R},\:(\delta^i_j-g^i_j(p))X^j=0\}.$$
The key observation is that $X^iv_i$ vanishes at $p$ if and only if $g^i_j(p)X^j=0.$ Using the association the standard commutation relations on homogeneous spaces\cite{coq}
\begin{eqnarray*}
[h_p,m_p]&\subset& m_p\\
\: [h_p,h_p]&\subset& h_p
\end{eqnarray*}
give the identity
\begin{equation}
(\delta^m_n-g^m_n)C^i_{nk}g^k_j=(\delta^m_n-g^m_n)g^i_kC^k_{mj}.
\end{equation}

The dual of the algebra (\ref{algebra}) is 
\begin{equation}
d\omega^i=-\frac{1}{2}(2\delta^i_m-g^i_m)C^m_{kl}\omega^k\wedge\omega^l
\end{equation}
and we obtain the formulas for the connection
\begin{eqnarray}
\nabla_{v_i}v_j&=&\Gamma^k_{ij}v_k\\
\Gamma^k_{ij}&=&\frac{1}{2}K^{kl}(C^n_{ij}g_{ln}-C^n_{li}g_{jn}+C^n_{jl}g_{il})\label{con}
\end{eqnarray}
and the curvature
\begin{eqnarray}
R(v_i,v_j)v_k&=&R^l_{kij}v_l\\
R^l_{kij}&=&C^m_{ij}\Gamma^l_{mk}+C^m_{ik}\Gamma^l_{jm}-C^m_{jk}\Gamma^l_{im}
+\Gamma^m_{jk}\Gamma^l_{mi}-\Gamma^m_{ik}\Gamma^l_{mj}\label{cur}.
\end{eqnarray}
The formula for the connection can be checked using the formula\cite{ber}
$$
2(\nabla_X Y\cdot Z )=X(Y\cdot Z)-Z(X\cdot Y)+Y(Z\cdot X)
+[X,Y]\cdot Z - [Y,Z]\cdot X -[Z,X]\cdot Y
$$
where $\cdot$ means contraction with respect to the metric $g.$ 
The Laplacian on zero-forms is actually the Casimir
\begin{equation}
\triangle_0=K^{ij}v_iv_j.\label{lap}
\end{equation}
One can also check that these manifolds are of constant scalar curvature
by calculating
$$
R=R^i_{jkl}g^j_ig^{kl}
$$
and checking that 
$$
L_{v_i}R=0
$$
(Appendix \ref{appb}).
 
We make few comments on the connection and the curvature.
First we have on $TM$ the usual Levi-Civita connection  
(\ref{con}) which is denoted as 
\begin{equation}
D=D_{L-C}=\delta^{\alpha}_{\beta}d+dx^{\gamma}\Gamma^{\alpha}_{\gamma\beta}.
\end{equation}
Since
\begin{eqnarray*}
\Gamma^{\gamma}_{\alpha\beta}&=&(\partial_{\alpha}\omega_{\beta}^j)v_j^{\gamma}+
 \omega_{\alpha}^i\omega_{\beta}^j\Gamma^k_{ij}v_k^{\gamma}\\
&=&(\partial_{\alpha}\omega_{\beta}^j)v_j^{\gamma}+\frac{1}{2}
 \omega_{\alpha}^i\omega_{\beta}^jC^k_{ij}v_k^{\gamma},
\end{eqnarray*}
we can write
\begin{eqnarray}
D^{\mu}_{\nu}&=&v_i^{\mu}(\delta^i_jd+\frac{1}{2}
\omega^kC^i_{kj})\omega^j_ {\nu}\\
&=&v_i^{\mu}\tilde{D}^i_j\omega^j_{\nu},
\end{eqnarray}
where $\tilde{D}$ is a differential operator.
There is also another interesting connection on
$TM$, which we call the Maurer-Cartan connection:
\begin{equation}
(D_{M-C})^{\mu}_{\nu}=v_i^{\mu}(\tilde{D}_{M-C})^i_j\omega^j_{\nu}\label{mau}
\end{equation}
where
\begin{equation} 
(\tilde{D}_{M-C})^i_j=\delta^i_jd+\omega^kC^i_{kj}.\label{mautilde}
\end{equation}
The differential operator associated with it has the property
\begin{equation}
(\tilde{D}_{M-C})^i_jg^j_k=g^i_j(\tilde{D}_{M-C})^j_k.
\end{equation}
The fact that it really is a connection is checked by calculating
the difference of it and the Levi-Civita connection which yields
the tensor
\begin{equation}
C_{\kappa\mu\nu}=\omega_{\mu}^i\omega_{\nu}^jC^k_{ij}K_{kl}\omega_{\kappa}^l.
\label{tensor}
\end{equation}
Then one can check easily that the axioms of a connection are satisfied
except the torsion free axiom\cite{wal}. The definition of
the tensor (\ref{tensor}) is invariant under rotations that take $v_i$ into their
linear combinations but may depend on the particular way
that $M$ is quotiented $M=G/H$. In addition it
is antisymmetric in all the three indices. Consequently, on any two dimensional homogeneous manifold $D$ and $D_{M-C}$ coincide.

Then we calculate the curvature of the Maurer-Cartan tensor, but
first the Levi-Civita curvature:
$$
D^2=\frac{1}{2}R^{\alpha}_{\beta\gamma\delta}dx^{\gamma}\wedge
dx^{\delta}
$$
where the curvature tensor is easily expressed in terms of
the tensor (\ref{cur}) as $R^{\alpha}_{\beta\gamma\delta}=v_i^{\alpha}R^i_{jkl}\omega^j_{\beta}
\omega^k_{\gamma}\omega^l_{\delta}.$
One can calculate the Maurer-Cartan curvature 
\begin{equation}F^{\alpha}_{\beta\gamma\delta}=v_i^{\alpha}F^i_{jkl}\omega^j_{\beta}
\omega^k_{\gamma}\omega^l_{\delta}\end{equation}
similarly:
\begin{eqnarray*}
D_{M-C}^2&=&v_i^{\mu}\tilde{D}_{M-C}g^i_j\tilde{D}_{M-C}\omega^j_{\nu}\\
&=&v_i^{\mu}\tilde{D}_{M-C}^2\omega^j_{\nu}\\
&=&\frac{1}{2}F^i_{jkl}v_i^{\mu}\omega^j_{\nu}\omega^k_{\kappa}\omega^ 
l_{\lambda}dx^{\kappa}\wedge dx^{\lambda}
\end{eqnarray*}
where
\begin{equation}
F^i_{jkl}=C^i_{jn}(\delta^n_m-g^n_m)C^m_{kl}.
\end{equation}
In the Lie group case $g^n_m=\delta^n_m$ and $F=0$ which
is very natural because on a Lie group
$$
D_{M-C}=g^{-1}dg=d+\Omega
$$
where $\Omega=g^{-1}dg$ is the Maurer-Cartan form ($g$ is now
exceptionally a group element $g\in G$, $g=e^{\theta^iT_i}$ where
$\theta$ are the exponential coordinates on the Lie group and 
$T_i$ are the generators of the Lie algebra) and the
connection is zero-curvature by the Maurer-Cartan equation\cite{sat}
$$
D_{M-C}^2=d\Omega+\Omega\wedge\Omega=g^{-1}d^2g=0.
$$

\Section{Formulas on the cotangent bundle}
\label{sec:pha}
We introduce some formulas and a one-parameter family of invariant
metrics on the cotangent bundle of homogeneous manifolds.

The phase space of the geodesic motion on a manifold is the
cotangent bundle. If $x^{\mu},p_{\mu}$ are the standard coordinates
on it $$\left.p_{\mu}dx^{\mu}\right|_p\in T^*M,\:\:p\in M $$
the standard symplectic potential is
$$\theta=p_{\mu}dx^{\mu}$$
and the Hamiltonian of the geodesic motion is
$$H=\frac{1}{2}g^{\mu\nu}p_{\mu}p_{\nu}.$$

The $G$ action on $M$ has a Hamiltonian lift\cite{2qc}, the Hamiltonians
that generate it are
\begin{equation}
I_i=v_i^{\mu}p_{\mu}
\end{equation} and they satisfy the Poisson brackets
\begin{equation}
\{I_i,I_j\}=C^k_{ij}I_k.\end{equation}
The associated Hamiltonian vector fields are
\begin{equation}
v_i^H=\{I_i,\:\:\}=v_i-\frac{\partial I_i}{\partial x^{\mu}}\frac{\partial}
{\partial p_{\mu}}.
\end{equation}
The symplectic form\cite{2qc} can be written as
\begin{equation}
\omega=d\theta=d(I_i\omega^i)=dI_i\wedge\omega^i-\frac{1}{2}I_iC^i_{jk}\omega^j\wedge \omega^k
\end{equation}
where $\omega^i$ is actually the pull-back of the
familiar form on $M$ onto the cotangent bundle by the canonical projection.

The dual vector fields to the forms $\omega^i$, $dI_i$ are
\begin{eqnarray*}
u_i&=&v_i^H-C^k_{ij}I_k w^j\\
w^i&=&\omega^i_{\mu}\frac{\partial}{\partial p_{\mu}},
\end{eqnarray*}
and the word ``dual'' means that the identity tensor in the fiber
of the tangent bundle can be written as
\begin{equation}
\mbox{\bf 1}_{2D}=u_i\otimes\omega^i+w^i\otimes dI_i=v_i^H\otimes\omega^i+w^i
\otimes(dI_i+C^k_{ij}I_k\omega^j).
\end{equation}
However, we prefer the vector fields $v_i^H$, $w^i$ whose contractions
are
\begin{eqnarray}
v_i^H\cdot dI_j&=&C^k_{ij}I_k\\
v_i^H\cdot \omega^j&=&g_i^j\\
w^i\cdot dI_j&=&g^i_j\\
w^i\cdot\omega^j&=&0
\end{eqnarray}
and commutators
\begin{eqnarray*}
[w^i,w^j]&=&0\\ \:
[w^i,v_i^H]&=&C^i_{jk}w^k\\ \:
[v_i^H,v_j^H]&=&C^i_{jk}v_i^H.
\end{eqnarray*}

The inverse of the symplectic form can be written using
these vector fields as
\begin{equation}
\omega^{-1}=v_i^H\wedge w^i+\frac{1}{2}I_iC^i_{jk}w^j\wedge w^k.
\end{equation}

Finally, we describe a natural invariant metric on the cotangent bundle.
Observing that
\begin{eqnarray*}
L_{v_i^H}\omega_j&=&C^k_{ij}\omega_k\\
L_{v_i^H}dI_j&=&C^k_{ij}dI_k
\end{eqnarray*}
the following tensors are invariant under the action of the group $G$ 
because of the Casimir structure
\begin{eqnarray}
g_1&=&K_{ij}\omega^i\otimes\omega^j\\
g_2&=&K^{ij}dI_i\otimes dI_j.
\end{eqnarray}
From these two one can combine a metric
\begin{equation}
g=\alpha g_1+\beta g_2 \label{phasespacemetric}
\end{equation}
provided that $\alpha$ and $\beta$ are positive.\\
{\it Proof:} The combination 
is positive-semidefinite because $K_{ij}$ is positive-definite.
One needs to check that it
is nondegenerate which can be done by calculating the determinant of it
$$
\det g=\beta\det g_1^{\mu\nu}\det(\alpha g^1_{\mu\nu}+\beta(K^{ij}-g^{ij})
\partial_{\mu}I_i\partial_{\nu}I_j)
$$
using  the block matrix formula and proving that it is non-zero, which is not difficult because the last term in the latter determinant is positive-semidefinite.$\Box$

The invariance property of the metric can be written explicitly as
$$L_{v_i^H}g=0,$$
see also Ref. \cite{2qc} for a different metric.

\Section{Bound for the geodesic action}\label{appa}
We derive a bound for the geodesic action
\begin{equation}
S=\int_0^{\beta}g_{\mu\nu}\dot{x}^{\mu}\dot{x}^{\nu}
\end{equation}
starting from the expression\cite{mor}
\begin{equation}
S(t)=\int g_{\mu\nu}(\dot{x}^{\mu}+tg^{\mu\kappa}\theta_{\kappa})
(\dot{x}^{\nu}+tg^{\nu\lambda}\theta_{\lambda})\geq 0
\end{equation}
where $\theta$ is an arbitrary one-form. The inequality holds for all $t$, especially
at the minimum
\begin{equation}
t=-\frac{\int\theta_{\mu}\dot{x}^{\mu}}{\int g^{\mu\nu}\theta_{\mu}
\theta_{\nu}}
\end{equation}
where it gives the bound
\begin{equation}
\int g_{\mu\nu}\dot{x}^{\mu}\dot{x}^{\nu}\geq
\frac{(\int\theta_{\mu}\dot{x}^{\mu})^2}{\int g^{\mu\nu}\theta_{\mu}
\theta_{\nu}}.
\end{equation}

For BRST quantization of the winding
number action that appears in the bound
see Ref. \cite{bau}.

\Section{Calculation of the scalar curvature and a formula}\label{appb}
We calculate the scalar curvature on homogeneous manifolds, but first
we calculate
\begin{equation}
g^a_ng_b^kg_c^ig_d^jR^n_{kij}\label{temp}
\end{equation}
using
\begin{enumerate}
\item
the missing identity
\begin{equation}
\Gamma^k_{ij}g_k^lg^i_a=\Gamma^k_{aj}g_k^l,\label{temp1}
\end{equation}
\item
the torsion
\begin{equation}
\Gamma^k_{ij}-\Gamma^k_{ji}=g^k_lC^l_{ij}\label{temp2}
\end{equation}
\item 
and the following property of the connection coefficient
\begin{equation}
g_a^ig_b^j\Gamma^k_{ij}g^c_k=
\frac{1}{2}g_a^ig_b^jC^k_{ij}g^c_k\label{temp3}.
\end{equation}
\end{enumerate}
The point is to use (\ref{temp1}) to eliminate the $\Gamma$:s using
(\ref{temp3}) when one substitutes the expression (\ref{cur}) in (\ref{temp}).
Another way is just hard work but it requires perhaps the introduction
of Feynman rules for the terms.
The result is
\begin{eqnarray}
&=&g^a_ng^k_bg^i_cg^j_d(-C^m_{ij}C^n_{mk}+\frac{1}{2}C^m_{ij}g_m^pC^n_{pk}\\
&+&\frac{1}{4}(C^m_{ik}g_m^pC^n_{pj}-C^m_{jk}g_m^pC^n_{pi}))
\end{eqnarray}
which is also useful in calculating the equation (\ref{hjelppia}). 
The scalar curvature is
\begin{eqnarray}
R&=&R^i_{jkl}g^i_jg^{kl}\\
&=&K^{ij}C^m_{ik}g^k_lC^l_{jm}-\frac{3}{4}K^{ij}C^m_{ik}g^k_lC^l_{jn}g^n_m\\
&=&K^{ij}tr(C_igC_j)-\frac{3}{4}K^{ij}tr(C_igC_jg).
\end{eqnarray}
Using the equation
\begin{equation}
L_{v_a}g^k_l= [g, C_a ]^k_l=g^k_pC^p_{al}-C^k_{ap}g^p_l
\end{equation}
it is easy to see that $L_{v_a}R=0$.
If one puts $g^i_j=\delta^i_j$ one recovers the Lie group formula
\begin{equation}
=\frac{1}{4}K^{ij}tr\:C_iC_j=\frac{D}{4},
\end{equation}
where $D$ is the dimension of the Lie group.

Finally we mention a formula. We define first 
\begin{eqnarray}
D_{-}=D_t-\frac{1}{2}\omega^i(\dot{x})
C^l_{ik}v_l^{\lambda}\omega^k_{\kappa}=v_i^{\lambda}\partial_t
\omega^i_{\kappa}\\
D_{+}=D_t+\frac{1}{2}\omega^i(\dot{x})C^l_{ik}v_l^{\lambda}
\omega^k_{\kappa}=D_t^{M-C}
\end{eqnarray}
where the operators are considered on classical trajectories of the geodesic
motion which means that $\omega^i(\dot{x})$ is constant.
The formula reads
\begin{equation}
D_{+}D_{-}=D_{-}D_{+}=D_t^2+R^{\alpha}_{\beta\gamma\delta}\dot{x}^{\beta}
\dot{x}^{\gamma}+F^{\alpha}_{\beta\gamma\delta}\dot{x}^{\beta}\dot{x}^{\gamma}
\label{hjelppia}
\end{equation}
where $F$ is the curvature of the Maurer-Cartan connection. In Lie
group case it reduces to the factorization of the geodesic deviation operator
$$\left.g^{\mu\kappa}\frac{\delta S}{\delta x^{\kappa}\delta x^{\nu}}\right|_{\delta S=0}
=D_{+}D_{-}=D_{-}D_{+}=D_t^2+R^{\alpha}_{\beta\gamma\delta}\dot{x}^{\beta}
\dot{x}^{\gamma}$$
where $S=\frac{1}{2}\int g_{\mu\nu}\dot{x}^{\mu}\dot{x}^{\nu}.$

\end{document}